# Influence of simultaneous doping of Sb & Pb on phase formation, superconducting and microstructural characteristics of $HgBa_2Ca_2Cu_3O_{8+\delta}$

**Rajiv Giri, R. S. Tiwari and O. N. Srivastava**

Department of Physics, Banaras Hindu University, Varanasi, INDIA

**Abstract**

We report systematic studies of structural, microstructural and transport properties of $(Hg_{0.80}Sb_{0.2-x}Pb_x)Ba_2Ca_2Cu_3O_{8+\delta}$ (where x = 0.0, 0.05, 0.1, 0.15, 0.2) compounds. Bulk polycrystalline samples have been prepared by two-step solid-state reaction route at ambient pressure. It has been observed that simultaneous substitution of Sb and Pb at Hg site in oxygen deficient $HgO_\delta$ layer of $HgBa_2Ca_2Cu_3O_{8+\delta}$ cuprate high-$T_c$ superconductor leads to the formation of Hg-1223 as the dominant phase. Microstructural investigations of the as grown samples employing scanning electron microscopy reveal single crystal like large grains embodying spiral like features. Superconducting properties particularly transport current density ($J_{ct}$) have been found to be sensitive to these microstructural features. As for example $(Hg_{0.80}Sb_{0.05}Pb_{0.15})Ba_2Ca_2Cu_3O_{8+\delta}$ compound which exhibits single crystal like large grains (~ 50 μm) and appears to result through spiral growth mechanism, shows highest $J_{ct}$ (~ $1.85 \times 10^3$ A/cm$^2$) at 77K. A possible mechanism for the generation of spiral like features and correlation between microstructural features and superconducting properties have been put forward.

Email : rajivgiri.phy@gmail.com, hepons@yahoo.com

## Introduction

Hg-based cuprate superconductors possess potential for practical applications due to their high critical temperature up to ~135 K (highest $T_c$ among those of known superconductors) [1-3]. These Hg-cuprates, however, show the weak pinning behavior and are difficult to synthesise. One of the implications of the extremely short coherence length in cuprates, typically a few angstrom, is that superconducting properties especially critical current density of these materials are sensitive to their underlying microstructure. A significant improvement in superconducting properties of both bulk and film forms of superconductors has been brought by several workers through understanding and control of their microstructural characteristics for Y and Bi based superconductors [4-12]. Among various kid of defects line defects such as screw dislocations, capable of pinning a vortex along a substantial part of its length, may extend the temperature and magnetic field range in which these high-$T_c$ cuprates may be utilized. The dislocations in favorable condition may lead to spiral growth where each grain or column grows by atoms to a spirally expanding ramp on the top surface of the grain [13]. This is exactly the classical growth mechanism proposed by Frank [14] and developed theoretically by Buston [15] and observed experimentally for the first time by Verma [16].

Although optimization of $T_c$ has been carried out by several workers [17-20], the pinning mechanism in Hg family of high temperature superconductor is not yet clear. Recently effect of Sb doping in Hg-1223 has been studied by several workers [17, 18]. They have found that small addition of Sb in Hg-1223 can improve the intergrain critical current density. However, Pb doping has been found to promote the formation of nearly single-phase of Hg-1223 compound and enhancement of irreversible field [19, 20, 21]. Li et al [22] have studied and optimized the synthesis and processing parameters that affect the phase purity and grain growth of Pb-doped $HgBa_2Ca_2Cu_3O_{8+\delta}$ cuprate superconductor. Recently Passos et al have discussed the effect of oxygen content on the pinning energy and critical current density in Hg-1223 [23].

Above mentioned individual doping of Sb and Pb in Hg-1223 leads to different specific effects on synthesis, structural/ microstructural and superconducting properties. However, to the best of our knowledge, simultaneous doping of Sb & Pb in Hg-1223 has not been carried out so far. In the present study we, therefore, have investigated the influence of simultaneous doping of Sb & Pb on the phase formation, superconducting properties and microstructural characteristics in Hg-1223 high-$T_c$ cuprate superconductor. It has been found that optimal doping of Sb and Pb in $HgBa_2Ca_2Cu_3O_{8+\delta}$ results in the enhancement of transport critical current density ($J_{ct}$). The interrelationship between observed microstructural features and $J_{ct}$ has been pointed out. A curious feature regarding the microstructure of $(Hg_{0.8}Sb_{0.05}Pb_{0.15})Ba_2Ca_2Cu_3O_{8+\delta}$ is the occurrence of spiral features. These are taken to originate due to the creation of screw dislocation arising out of the misfit between step heights ('c' parameters) of Hg(Sb,Pb)-1223 and Hb(Sb, Pb) -1234 regions.

**Experimental**

In the present study the synthesis of Sb and Pb doped $HgBa_2Ca_2Cu_3O_{8+\delta}$ high temperature cuprate superconductor has been carried out by two-step solid-state reaction method. In the first step highly pure (3N) powders of $Ba(NO_3)_2$, CaO and CuO were mixed in the stoichiometric ratio ( Ba/Ca/Cu=2/2.2/3.2) and ground in an agate mortar. The resultant mixture was heated in flowing oxygen with three intermediate grindings at $925^0$C for 48 h. In the second step the Ba-Ca-Cu-O precursor was mixed with HgO, $Sb_2O_3$ and PbO to obtain the nominal compositions $(Hg_{0.80}Sb_{0.2-x}Pb_x)Ba_2Ca_2Cu_3O_{8+\delta}$ (where x =0.0, 0.05, 0.1, 0.15, 0.2). All the intermediate grindings and final mixing/ grinding have been carried out in a glove box containing $P_2O_5$, NaOH and nitrogen atmosphere. After final grinding the Hg(Sb,Pb)-Ba-Ca-Cu-O powder was pelletized at a pressure of 3.5 tons/cm$^2$. The pellets were put in a platinum box, which was then put into a quartz tube. The quartz tube was evacuated up to $10^{-5}$ torr and sealed. For safety purposes the sealed tube was encased in a steel tube filled with sand and tightly closed. Thereafter, steel tube was inserted into a programmable Heraeus tube type furnace. The temperature of the furnace was raised at a rate of $200^0$C/h up to $600^0$C and thereafter at a rate of $100^0$C/h up to $840^0$C and held at this temperature for 12 h. Finally the furnace was cooled at the rate of $100^0$C/h to room temperature. All the samples in the present investigations were subjected to gross structural characterization by X-ray diffraction (XRD, Philips PW 1710, $CuK_\alpha$), electrical transport characterization by four-probe technique (Keithley resistivity-Hall setup), surface morphological characterization by scanning electron microscope (SEM, Phillips XL20) and elemental composition investigation by energy dispersive analysis of X-ray (EDAX;

PV9761/75) microanalysis system, attached to the TECNAI $20G^2$ transmission electron microscope.

**Results and Discussion**

As grown samples of $(Hg_{0.80}Sb_{0.2-x}Pb_x)Ba_2Ca_2Cu_3O_{8+\delta}$ having various Sb and Pb concentrations were subjected to gross structural characterization employing X-ray diffraction (XRD) technique. The XRD data collected from various samples show that all the samples are polycrystalline and correspond to tetragonal Hg(Sb/Pb)-1223 superconducting phase. The XRD pattern also shows low $T_c$ Hg-1234 phase and some impurity phases with vanishingly small concentrations. The representative XRD patterns are shown in Fig.1. The lattice parameters have been estimated using d-values and (hkl) reflections of the observed X-ray diffraction and these are shown in table 1.

The stoichiometry corresponding to various compositions was verified by EDAX, which reveals that the estimated stoichiometry is found to be conformity with the envisaged stoichiometry.

The variation of resistance with temperature of the as synthesized $(HgSb_{0.2-x}Pb_x)$-$Ba_2Ca_2Cu_3O_{8+\delta}$ HTS samples was measured by the standard four-probe technique. The normal state resistance of all the samples show metal like behaviour with respect to temperature. The representative R-T behavior of the as grown $(Hg_{0.80}Sb_{0.2-x}Pb_x)Ba_2Ca_2Cu_3O_{8+\delta}$ (where $0\leq x \leq 0.2$) samples are shown in Fig.2. It may be noticed that all the samples, which are subjected to $T_c$ measurement, have similar geometry. From our investigation of R-T behavior, it may be inferred that normal state resistance of the optimum sample i.e. $(Hg_{0.80}Sb_{0.05}Pb_{0.15})$-1223 is lower than other ones, which reflects the presence of optimum carrier concentration in this compound. The values of critical transition temperature ($T_c$) for as grown $(Hg_{0.80}Sb_{0.20})$-1223, $(Hg_{0.80}Sb_{0.15}Pb_{0.05})$-1223, $(Hg_{0.80}Sb_{0.10}Pb_{0.10})$-1223, $(Hg_{0.80}Sb_{0.05}Pb_{0.15})$-1223 and $(Hg_{0.80}Pb_{0.20})$-1223 HTS compounds are 126K, 127K, 129K, 133K and 130K respectively within the accuracy of ± 0.5K.

The transport critical current density ($J_{ct}$) of all the samples have been measured by the four-probe technique using the criterion of 1μV/cm. The variation of $J_{ct}$ with temperature is shown in Fig.3. The highest $J_{ct}$ value ~$1.85\times10^3 A/cm^2$ at 77K has been found for $(Hg_{0.80}Sb_{0.05}Pb_{0.15})Ba_2Ca_2Cu_3O_{8+\delta}$ compound. It may be pointed out that the values of $T_c$ and $J_{ct}$ are sensitive to the doping concentration of Sb and Pb in Hg-1223. The variation of $T_c$ and $J_{ct}$ as a function of dopant cationic concentration is shown in Fig.4. It is clear from the figure that optimum value of both, $T_c$ and $J_{ct}$ results for $(Hg_{0.80}Sb_{0.05}Pb_{0.15})Ba_2Ca_2Cu_3O_{8+\delta}$ compound.

The surface morphological features of all the as grown samples have been studied by scanning electron microscopy (SEM) employing secondary electrons. The details of the SEM morphology of various samples are shown in Fig.5. The morphological evaluation of $(Hg_{0.80}Sb_{0.2-x}Pb_x)Ba_2Ca_2Cu_3O_{8+\delta}$ samples bear some interesting features with varying concentration of Sb & Pb and a correlation with critical current density. Fig.5(a) shows the representative SEM micrograph corresponding to $(Hg_{0.80}Sb_{0.20})Ba_2Ca_2Cu_3O_{8+\delta}$ compound. SEM micrograph reveals porosity and random orientation of platelet like small grains. The measured $J_{ct}$ value for $(Hg_{0.80}Sb_{0.20})Ba_2Ca_2Cu_3O_{8+\delta}$ is ~730 A/cm$^2$ at 77K. The replacement of Sb by Pb in $(Hg_{0.80}Sb_{0.2-x}Pb_x)Ba_{2}Ca_2Cu_3O_{8+\delta}$ increases transport critical current density ($J_{ct}$). For example $J_{ct}$ value for $(Hg_{0.80}Sb_{0.15}Pb_{0.05})Ba_2Ca_2Cu_3O_{8+\delta}$ is ~865 A/cm$^2$ at 77K. The morphology of this compound as revealed by SEM exhibits large single crystal like grains (~15 μm) together with improved connectivity of grains [Fig.5(b)]. On further increasing Pb concentration upto 0.15 i.e. corresponding to composition $(Hg_{0.80}Sb_{0.05}Pb_{0.15})Ba_2Ca_2Cu_3O_{8+\delta}$, very large single crystal like platy grains (~50μm), which are densely packed and well connected, are found to be present. In order to show large number of grains for this composition, a low magnification micrograph has been given in Fig. 5(c). The $J_{ct}$ for this composition has been found to be ~1.85 x 10$^3$A/cm$^2$, which is the highest amongst the observed $J_{ct}$ values for all the samples with different compositions exhibiting different surface morphological details. It should be pointed out that the reason for highest $J_{ct}$ observed for this composition may lie in the fact that the grains are very large [for this grain size is ~50 μm in comparison to the grain size of $(Hg_{0.80}Sb_{0.20})Ba_2Ca_2Cu_3O_{8+\delta}$ compound (~10 μm)], which implies lowering down of grain boundary density. In the case of $(Hg_{0.80}Pb_{0.2})Ba_2Ca_2Cu_3O_{8+\delta}$, where Sb is completely replaced by Pb, large grains have been found to result [Fig 5(d)]. However, large grains for this compound are not so well connected as for $(Hg_{0.80}Sb_{0.05}Pb_{0.15})Ba_2Ca_2Cu_3O_{8+\delta}$. The $J_{ct}$ value for this phase is ~1.4x10$^3$ A/cm$^2$, which is lower than that of $(Hg_{0.80}Sb_{0.05}Pb_{0.15})$-1223.

It is worthwhile to discuss the comparative influence of Sb and Pb doping, on growth marphology and critical current density. It should be pointed out that Sb doping is known to restrain the $O_2$ and Hg in the grains of Hg-1223 [17]. The presence of Pb is known to lower the melting point [20]. One curious microstructural characteristics observed for the material $(Hg_{0.80}Sb_{0.05}Pb_{0.15})Ba_2Ca_2Cu_3O_{8+\delta}$ is the presence of large single crystal like grains which embody spiral features on their surfaces. These features are shown in Fig.5(c) and more clearly in Fig. 6(a&b). Significant work on spiral growth for several materials have been done in our laboratory [16, 24]. Based on this, we propose the following mechanism/ model for the occurrence of spirals and consequently the growth of large single crystal like grains for the present material $(Hg_{0.80}Sb_{0.05}Pb_{0.15})Ba_2Ca_2Cu_3O_{8+\delta}$. A spiral growth will result when suitable growth front leading to screw dislocation gets formed and the atoms/ molecules get incorporated at the step accompanying the screw dislocation. In the present case two different structures namely Hg(Sb, Pb):1223 and Hg(Sb, Pb):1234 generally get formed as evidenced through XRD characterization (Fig.1). The 'c' lattice parameter of the structures are different [for Hg(Sb, Pb):1223; 'c' parameter = 15.86 Å and for Hg(Sb, Pb):1234 'c' parameter is 19.02 Å; $\Delta c$=3.16 Å). This difference in 'c' will lead to formation of screw dislocations and the resulting step (whose height will be integral multiple of $\Delta c$ ) when the nuclei of Hg(Sb, Pb):1223 and Hg(Sb, Pb):1234 coalesce. The atoms/ molecules of Hg(Sb, Pb)Ba-Ca-Cu-O in the matrix (which due to lowering of melting point because of the presence of Pb, is expected to be in molten form) will diffuse to the exposed step which will start winding up. This will lead to the growth of a large single crystal like grains with spiral features on the surface of the grains. This is in keeping with the observations of microstructures [Fig. 6(a&b)].

It may be pointed out that based on the known characteristics, the vortex motion may arise due to self magnetic field. In the absence of flux pinning, the vortex flow will occur resulting in lower $J_{ct}$. However, defects like screw dislocations (which also leads to large grain growth through spiral mechanism) which are present will provide flux pinning centers leading to higher values of $J_{ct}$. The observed values of $J_{ct}$ in the present investigations are in keeping with the above said criterion. Thus, the highest current density $J_{ct}$ (~$1.85 \times 10^3$ A/cm$^2$) is obtained for $(Hg_{0.80}Sb_{0.05}Pb_{0.15})Ba_2Ca_2Cu_3O_{8+\delta}$ compound which exhibits the most prominent spiral growth (Figs. 6 a&b).

## Conclusions

In the present study, we have successfully synthesized $(Hg_{0.80}Sb_{0.2-x}Pb_x)Ba_2Ca_2Cu_3O_{8+\delta}$ high-$T_c$ superconducting compounds (x = 0.00, 0.05, 0.10, 0.15 and 0.20). We have investigated the effect of simultaneous doping of Sb & Pb at Hg site of $HgO_\delta$ layer in $HgBa_2Ca_2Cu_3O_{8+\delta}$ with special emphasis on correlation between superconducting properties and the observed microstructural features. The as synthesized compound correspond to Hg(Sb,Pb)-1223 as the majority phase and Hg(Sb,Pb)-1234 as the minority phase. The critical transition temperature ($T_c$) of the Sb & Pb doped Hg-1223 compounds range between 126-133K. The highest $T_c$ value 133K has been found for $(Hg_{0.80}Sb_{0.05}Pb_{0.15})Ba_2Ca_2Cu_3O_{8+\delta}$ compound. The superconducting properties particularly $J_{ct}$ has been found to be correlated with observed surface morphological features. These features are in the form of spirals, which have been proposed to originate due to the occurrence of screw dislocations when Hg(Sb,Pb)-1223 and Hg(Sb,Pb)-1234 crystal regions coalesce. The spiral growth leads to the formation of large single crystal like regions. This will thus result in lowering the grain boundary density. This effect together with the possible flux pinning arising from the screw dislocations can be taken to lead to high critical current density for the $(Hg_{0.80}Sb_{0.05}Pb_{0.15})Ba_2Ca_2Cu_3O_{8+\delta}$ compound.

## Acknowledgements

The authors are grateful to Prof. A.R. Verma, Prof. C.N.R. Rao, Prof. S.K. Joshi and Prof. A.K. Roychaudhary for fruitful discussion and suggestions. Financial supports from UGC, DST-UNANST and CSIR are gratefully acknowledged. One of the authors Rajiv Giri is grateful to CSIR New Delhi, Govt. of India for awarding SRF (Ext.) fellowship.

## Figure caption

**Fig.1** Representative X-ray diffraction patterns of $(Hg_{0.80}Sb_{0.2-x}Pb_x)Ba_2Ca_2Cu_3O_{8+\delta}$ with (a) x=0.00 (b)x=0.05 (c) x=0.10 (d) x=0.15 (e) x=0.20, as synthesized samples for various composition using $CuK_\alpha$ radiation.

**Fig.2** Resistance *vs*. temperature behaviour for as grown $(Hg_{0.80}Sb_{0.2-x}Pb_x)Ba_2Ca_2Cu_3O_{8+\delta}$ compounds.

**Fig.3** Transport critical current density as a function of temperature for as grown $(Hg_{0.80}Sb_{0.2-x}Pb_x)Ba_2Ca_2Cu_3O_{8+\delta}$ compounds.

**Fig.4** Variation of $T_c$ and $J_{ct}$ as a function of dopant cationic concentration of as grown $(Hg_{0.80}Sb_{0.2-x}Pb_x)Ba_2Ca_2Cu_3O_{8+\delta}$ compounds

**Fig.5.** SEM micrograph corresponding to **(a)** $(Hg_{0.80}Sb_{0.20})$-1223 compound showing random orientation of grains and their poor connectivity, **(b)** $(Hg_{0.80}Sb_{0.15}Pb_{0.05})$-1223 compound depicting better grain growth **(c)** $(Hg_{0.80}Sb_{0.05}Pb_{0.15})$-1223 compound showing very large single crystal like grains which are densely packed and well connected, **(d)** $(Hg_{0.80}Pb_{0.20})$-1223 compound showing large grains which are not so well connected as grains in figure 5(c).

**Fig.6** Grains of $(Hg_{0.80}Sb_{0.05}Pb_{0.15})$-1223 compound depicting **(a)** circular and **(b)** octagonal spiral like features.

**Table 1** Various compositions ($(Hg_{0.80}Sb_{0.2-x}Pb_x)Ba_2Ca_2Cu_3O_{8+\delta}$, transition temperature, lattice parameters and transport critical current density.

| **Compositions** | **$T_c$ (R=0) (K) ±0.5** | **Lattice Parameter a(Å) ±0.003** | **Lattice parameter c (Å) ±0.001** | **Transport critical current density ($J_{ct}$) in A/cm$^2$ at 100K** | **77K** | **50K** |
|---|---|---|---|---|---|---|
| $(Hg_{0.80}Sb_{0.20})$-1223 | 126 | 3.846 | 15.846 | 200 | 730 | 1180 |
| $(Hg_{0.80}Sb_{0.15}Pb_{0.05})$ -1223 | 127 | 3.845 | 15.854 | 240 | 865 | 1300 |
| $(Hg_{0.80}Sb_{0.10}Pb_{0.10})$ -1223 | 129 | 3.847 | 15.866 | 225 | 1075 | 1680 |
| $(Hg_{0.80}Sb_{0.05}Pb_{0.15})$ -1223 | 133 | 3.843 | 15.871 | 600 | 1850 | 2785 |
| $(Hg_{0.80}Pb_{0.20})$-1223 | 130 | 3.842 | 15.878 | 400 | 1420 | 2040 |

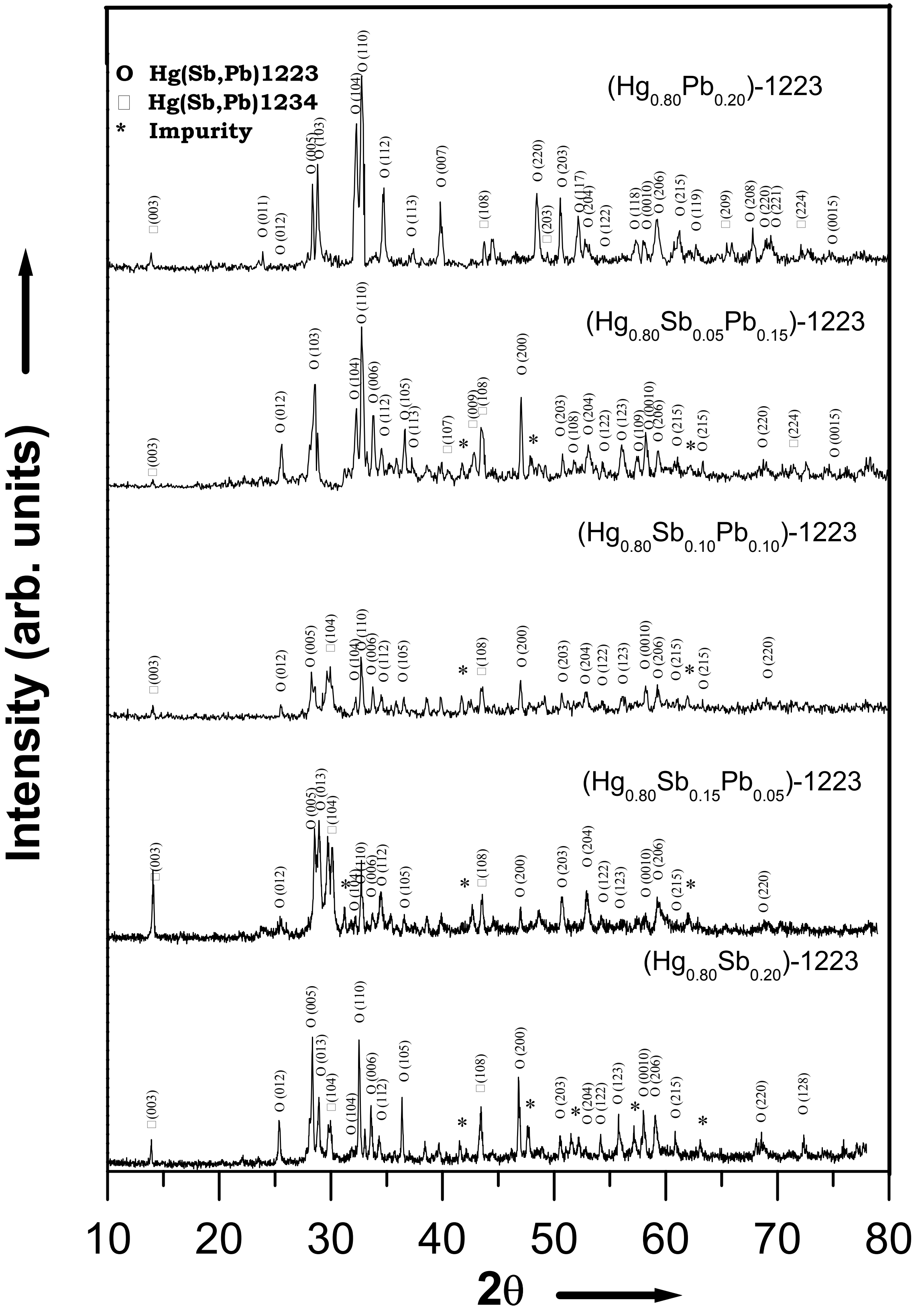

**Fig. 1**

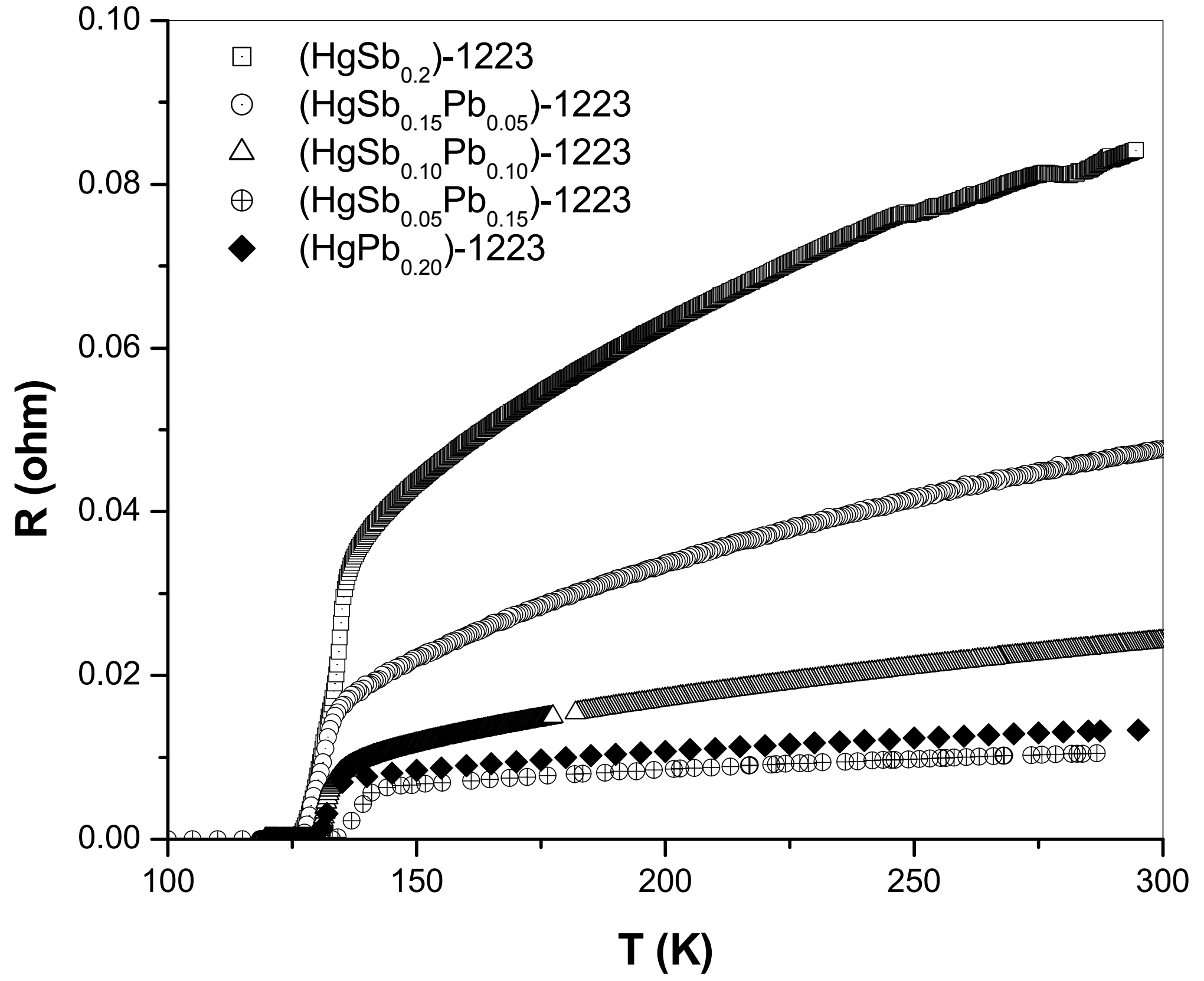

Fig. 2

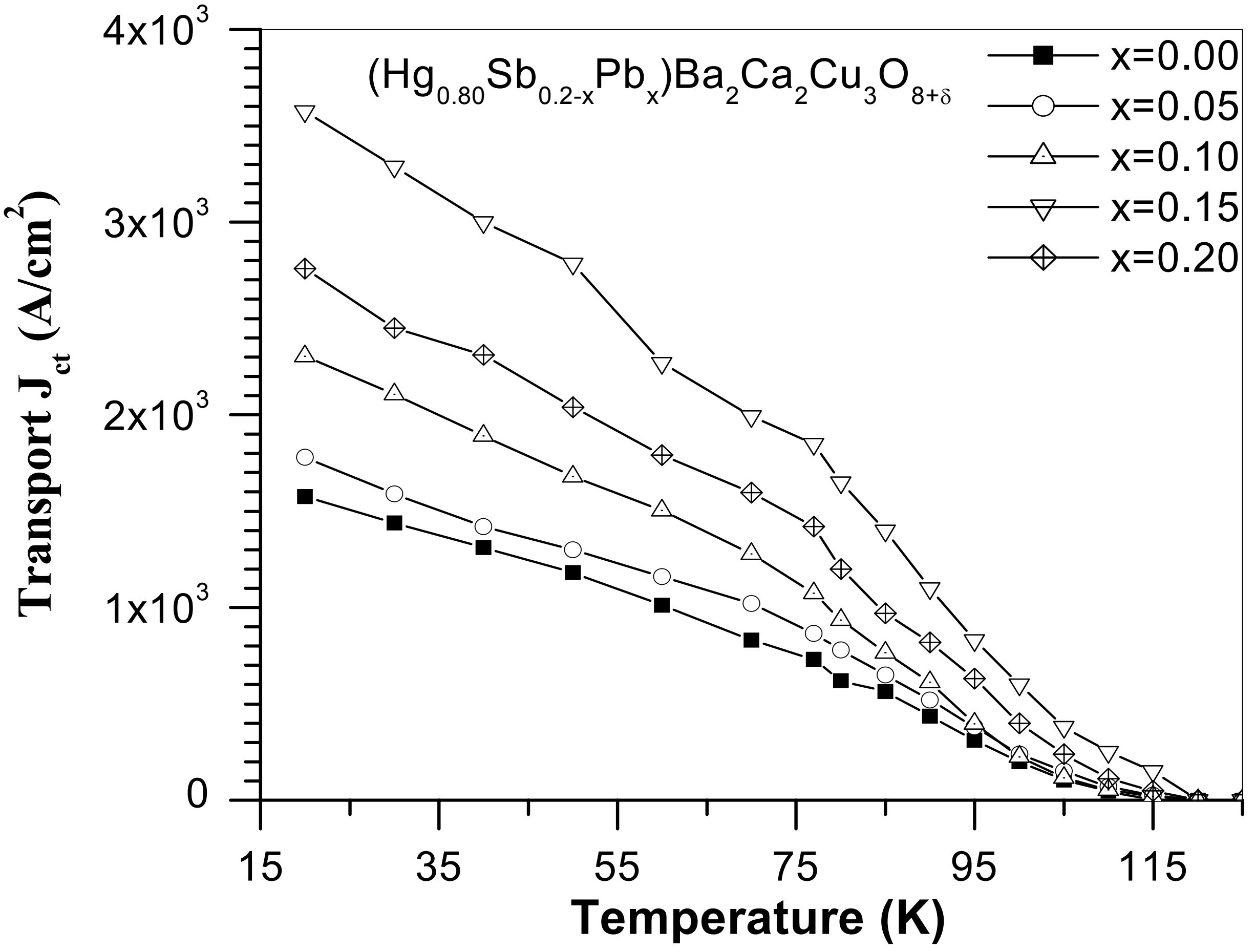

**Fig. 3**

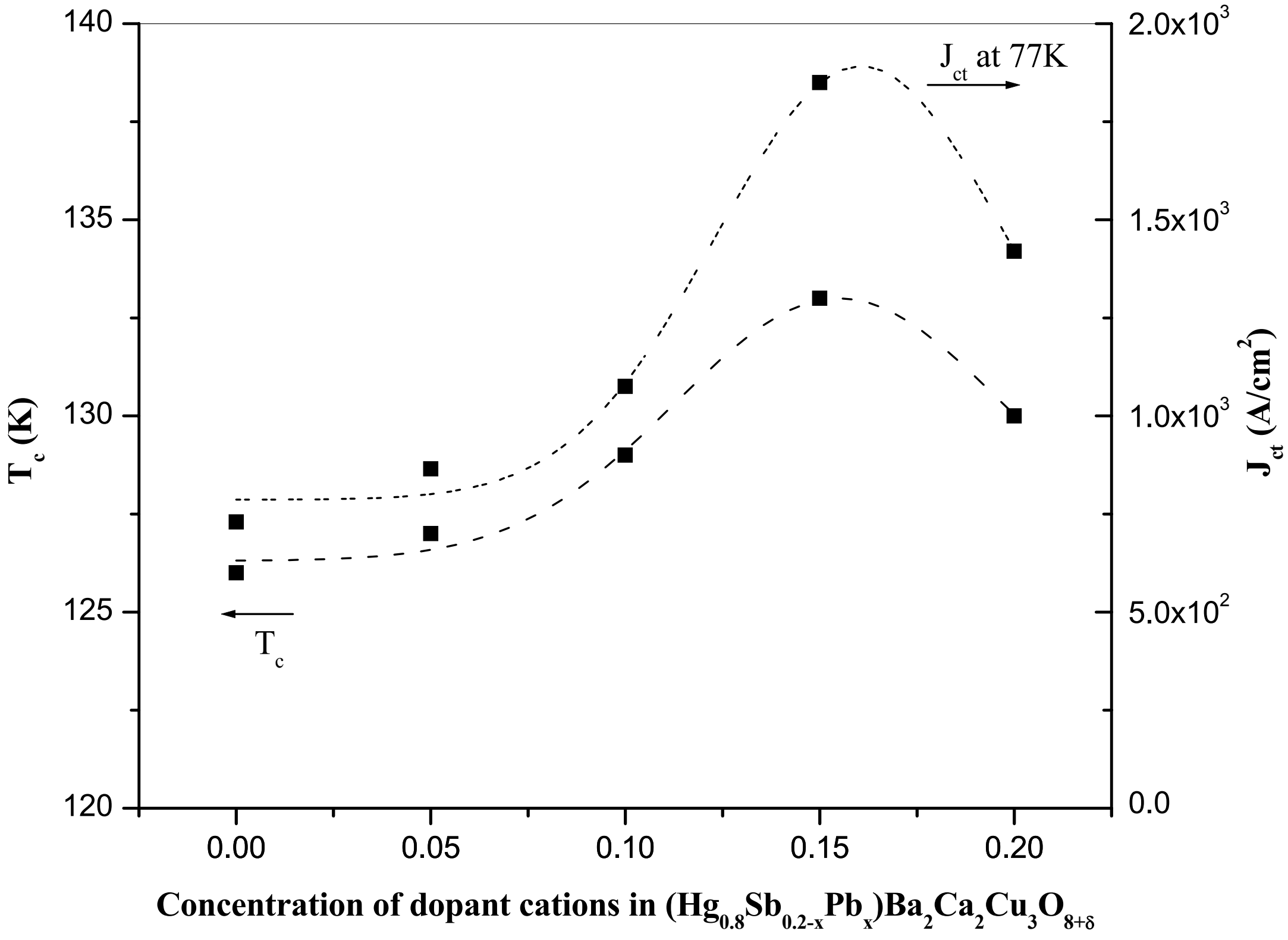

140
135
130
125
120
$T_c$ (K)
2.0x10$^3$
1.5x10$^3$
1.0x10$^3$
5.0x10$^2$
0.0
$J_{ct}$ (A/cm$^2$)
$J_{ct}$ at 77K
$T_c$
0.00
0.05
0.10
0.15
0.20
Concentration of dopant cations in $(Hg_{0.8}Sb_{0.2-x}Pb_x)Ba_2Ca_2Cu_3O_{8+\delta}$

**Fig. 4**

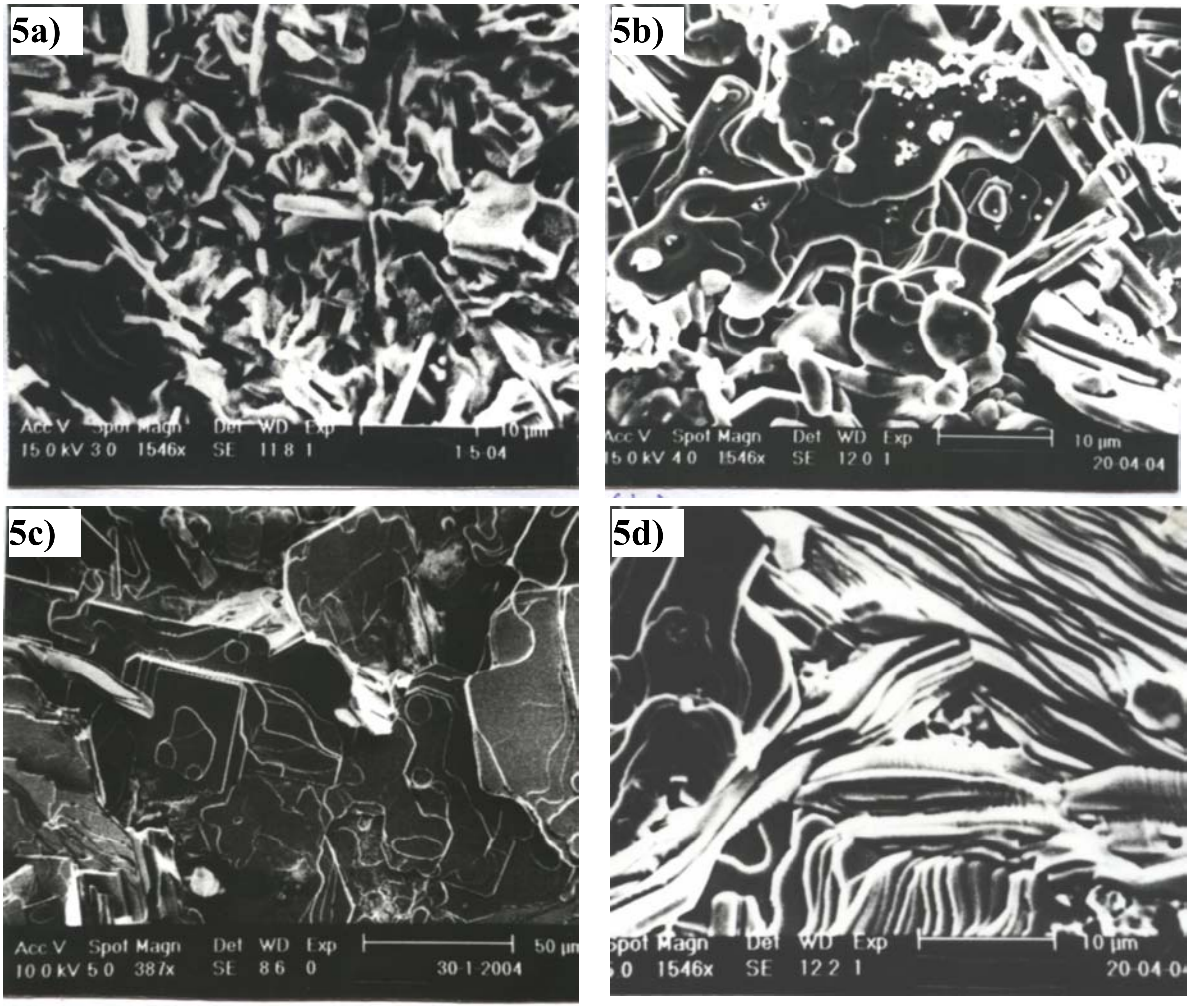
5a)
Acc V Spot Magn Det WD Exp 10 µm
15.0 kV 3.0 1546x SE 11.8 1 1-5-04
5b)
Acc V Spot Magn Det WD Exp 10 µm
15.0 kV 4.0 1546x SE 12.0 1 20-04-04
5c)
Acc V Spot Magn Det WD Exp 50 µm
10.0 kV 5.0 387x SE 8.6 0 30-1-2004
5d)
Spot Magn Det WD Exp 10 µm
0 1546x SE 12.2 1 20-04-0

**Fig. 5**

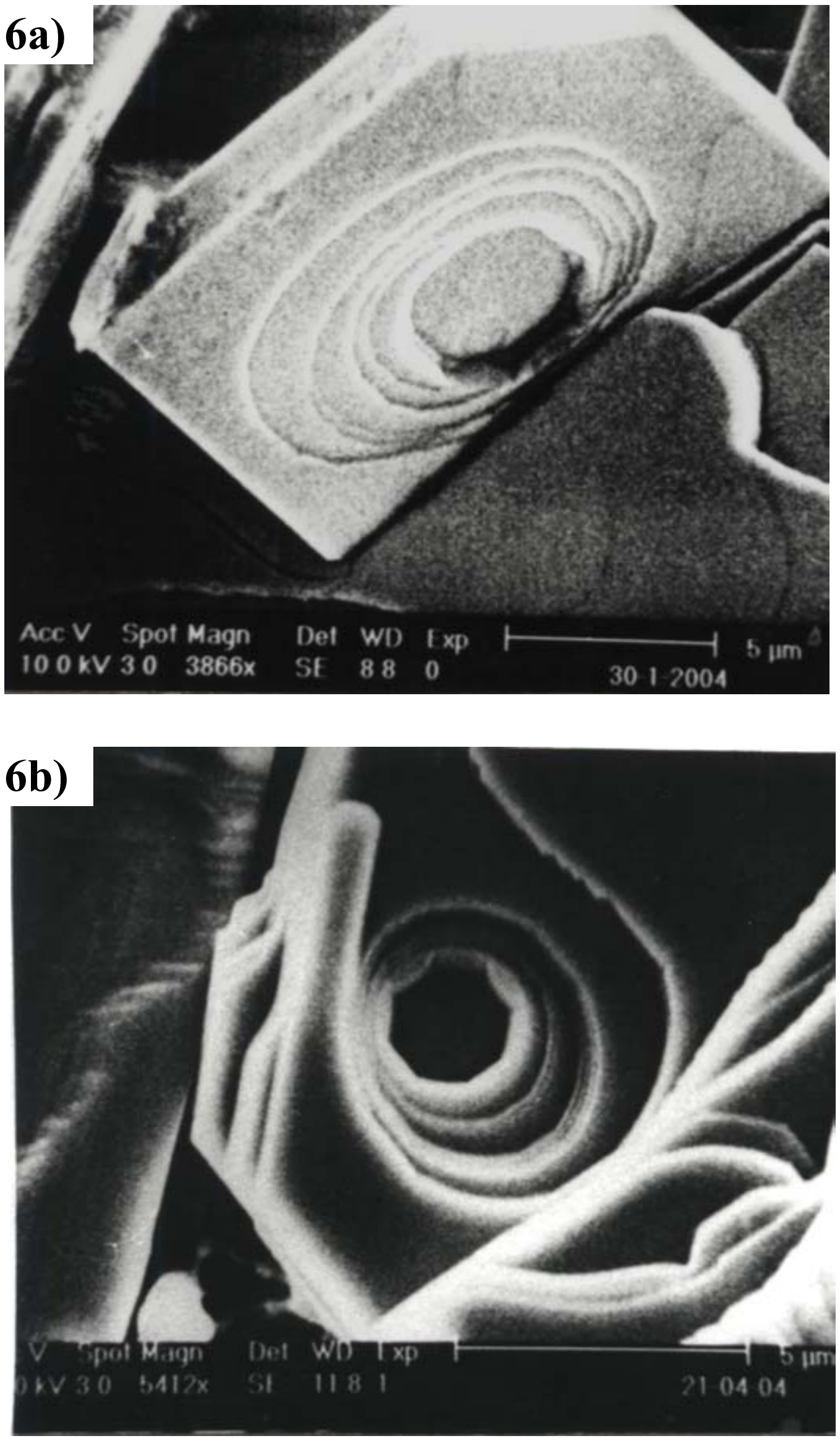
6a)
Acc V Spot Magn Det WD Exp 5 µm
10.0 kV 3.0 3866x SE 8.8 0 30-1-2004
6b)
V Spot Magn Det WD Exp 5 µm
0 kV 3.0 5412x SE 11.8 1 21-04-04

Fig. 6